# Highly responsive Y-Ba-Cu-O thin film THz detectors with picosecond time resolution

Petra Thoma, Juliane Raasch, Alexander Scheuring, Matthias Hofherr, Konstantin Il'in, Stefan Wünsch *Member IEEE*, Alexei Semenov, Heinz-Wilhelm Hübers, Vitali Judin, Anke-Susanne Müller, Nigel Smale, Jens Hänisch, Bernhard Holzapfel and Michael Siegel

*Abstract*—High-temperature superconducting $YBa_2Cu_3O_{7-\delta}$ (YBCO) thin-film detectors with improved responsivities were developed for fast time-domain measurements in the THz frequency range. YBCO thin films of $\approx 30$ nm thickness were patterned to micro- and nanobridges and embedded into planar log-spiral THz antennas. The YBCO thin-film detectors were characterized with continuous wave radiation at 0.65 THz. Responsivity values as high as 710 V/W were found for the YBCO nanobridges. Pulsed measurements in the THz frequency range were performed at the electron storage ring ANKA from the Karlsruhe Institute of Technology (KIT). Due to the high responsivities of the nanobridges no biasing was required for the detection of the coherent synchrotron radiation pulses achieving very good agreement between the measured pulse shapes and simulations.

## I. INTRODUCTION

ELECTRON storage rings generate broadband synchrotron radiation by deflecting accelerated electrons in bending magnets. As a direct consequence of the radio frequency power used for the acceleration of the electrons, the synchrotron radiation is not continuous but grouped in so-called bunches resulting in a pulsed emission. In particular in the terahertz (THz) frequency range, electron storage rings are powerful sources for the generation of brilliant ultra-short pulses [1]. By adapting the magnets to so-called "low-alpha optics", coherent synchrotron radiation (CSR) is generated with pulse widths (root mean square) in the single picosecond range [1] - [3]. To optimize the CSR generation, ultra-fast detectors which are able to monitor these picosecond pulses are required. Since the electron beam current, which is directly proportional to the emitted power, is adjustable over a very wide range going down to one stored electron [1], these ultra-fast detectors should also show a very broad dynamic range. Furthermore, large responsivity values are advantageous because then amplifiers which may influence the pulse shape are not required in the readout of these ultra-short pulses.

Existing direct THz detectors comprise room-temperature as well as cooled detector technologies. However, most of these detector technologies do not fulfill the requirement of picosecond response times. Room-temperature pyroelectric detectors [4] and Golay cells [5] show response times between 25 and 200 ms. Much shorter response times are achieved with cooled hot-electron bolometers based on NbN. However, these devices with response times of $\approx 50$ ps [6] - [8] are also too slow for the analysis of the temporal evolution of the CSR pulses. Schottky-barrier diodes show response times of a few picoseconds fulfilling the requirement of ultra-fast time resolution. First measurements of a zero-bias Schottky diode embedded in a quasi-optical detector design were reported [9]. A Schottky diode embedded in a broadband quasi-optical detector design shows responsivity values of 500 V/W [10].

Another promising candidate for ultra-fast response times is the high-temperature superconductor $YBa_2Cu_3O_{7-\delta}$ (YBCO). Response times of only a few picoseconds were measured from optical to infrared wavelengths using pump-probe and autocorrelation techniques [11] - [14]. We recently demonstrated picosecond response times in the THz frequency range [15]. Furthermore, a dynamic range larger than 30 dB was found [16]. Responsivity values of substrate-supported THz YBCO micrometer-sized detectors between 190 and 480 V/W were reported [17] - [19] while values up to 2900 V/W were achieved for suspended YBCO bridges [20]. This reveals the well-known inverse proportionality of the responsivity to the thermal conductance where the latter can be also reduced by decreasing the lateral dimensions of the detecting element to the sub-micrometer scale [17]. Thus, direct YBCO THz detectors are a promising technology for the analysis of ultra-short CSR pulses.

Manuscript received October 9, 2012. This work was supported by the German Federal Ministry of Education and Research (Grant Nos. 05K10VKD, 05K10KTC).

P. Thoma, J. Raasch, A. Scheuring, M. Hofherr, K. Il'in, S. Wünsch and M. Siegel are with the Institut für Mikro- und Nanoelektronische Systeme, Karlsruher Institut für Technologie, 76187 Karlsruhe, Germany (phone: +49 721 608 44447; e-mail: petra.probst@kit.edu).

A. Semenov is with the Institut für Planetenforschung, DLR e.V, 12489 Berlin, Germany (e-mail: Alexei.semenov@dlr.de).

H.-W. Hübers is with the Institut für Optik und Atomare Physik, Technische Universität Berlin, 10623 Berlin, Germany, and also with the Institut für Planetenforschung, DLR e.V., 12489 Berlin, Germany (e-mail: heinz-wilhelm.huebers@dlr.de).

V. Judin and A.-S. Müller are with the Laboratorium für Applikationen der Synchrotronstrahlung, Karlsruher Institut für Technologie, 76131 Karlsruhe, Germany (e-mail: Vitali.Judin@iss.fzk.de, Anke-Susanne.Mueller@iss.fzk.de).

N. Smale is with the ANKA Synchrotron Radiation Facility, Karlsruher Institut für Technologie, 76344 Eggenstein-Leopoldshafen, Germany (e-mail: nigel.smale@kit.edu).

J. Hänisch and B. Holzapfel are with the Leibniz-Institut für Festkörper- und Werkstoffforschung Dresden, IFW Dresden, 01069 Dresden, Germany (email: J.Haenisch@ifw-dresden.de, B.Holzapfel@ifw-dresden.de).

In this report, we demonstrate the application of sub-µm YBCO detectors to analyze THz radiation. In section II and III the development of our thin-film deposition and patterning technology towards submicrometer dimensions to optimize the responsivity of our YBCO detectors is described. Nanometer-sized bridges with a length of only 300 nm and a width of 900 nm were successfully fabricated (see section IV). In section V the characterization of the YBCO detectors at 0.65 THz with continuous wave radiation chopped at 15 Hz is shown. An increase of the responsivity of a factor $\approx 2$ compared to the micrometer-sized detecting elements was found. The highly responsive nanobridges were embedded in our YBCO detection system (section V.A) and used for the characterization of CSR at ANKA, the electron storage ring of the Karlsruhe Institute of Technology (KIT). Even without any bias a clear detector response was measured (section V.C) allowing to operate our YBCO thin-film nanometer-sized devices as zero-bias detectors.

## II. THIN FILM TECHNOLOGY

The YBCO thin films were fabricated using the pulsed-laser deposition (PLD) technique. In spite of good crystalline matching of $SrTiO_3$ or $LaAlO_3$ to YBCO (mismatch of 1.4% and 2% respectively) [21], we chose both-side polished R-plane sapphire as substrate for our YBCO thin films with a stronger lattice mismatch of 12% [22]. The reason for choosing sapphire are its low dielectric losses ($\varepsilon_r = 10.06$, $\tan\delta = 8.4 \cdot 10^{-6}$ at 77 K) [22] which are essential for the back illumination of our YBCO detectors at THz frequencies.

The sapphire substrates were cleaned using acetone and isopropanol in an ultrasonic bath. To ensure a good thermal contact during deposition the substrates were mounted on the heater with silver paste. After mounting the heated substrate holder was adjusted in on-axis position adjusting the heater surface with the substrate normal to the plasma-plume axis. The vacuum chamber was evacuated by a turbo pump to a base pressure below $1 \cdot 10^{-6}$ mbar.

For the deposition the substrates were heated up to 800°C with a ramp of 700°C/hour. Due to the crystalline mismatch between sapphire and YBCO [22], it is not possible to grow high-quality superconducting YBCO thin films directly on sapphire. Therefore, buffer layers to improve the matching between the crystalline structures have to be used in the deposition process. The PLD system is equipped with a carousel of six targets allowing in-situ fabrication of multilayer samples. A $CeO_2$ buffer layer of 8 nm thickness was deposited at a substrate temperature of 800°C and an oxygen pressure of $p_{O2} = 0.9$ mbar. Pulses from a KrF excimer laser (wavelength $\lambda = 248$ nm) were focused on the rotating $CeO_2$ target with a pulse energy density of $\approx 1.5$ J/cm$^2$ resulting in a deposition rate of 0.8 nm/s.

To further improve the crystalline matching with the YBCO film, an additional 25 nm thick buffer layer of $PrBa_2Cu_3O_{7-\delta}$ (PBCO) was deposited on top of the $CeO_2$ layer. For this, the deposition temperature and the laser energy density were kept constant, and the partial oxygen pressure was reduced to $p_{O2} = 0.7$ mbar. The PBCO deposition rate was 0.7 nm/s. The YBCO film with a thickness of $\approx 30$ nm was then deposited on top of the PBCO buffer layer at the same temperature, oxygen pressure and laser energy density. The YBCO deposition rate amounted to 0.8 nm/s.

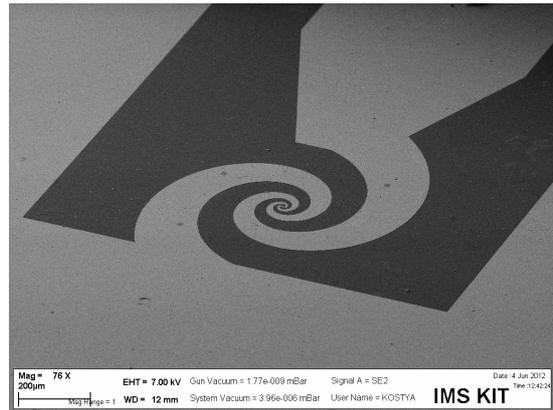

Fig. 1. SEM image of the planar log-spiral THz antenna. The bright part is the Au antenna whereas the black part shows the sapphire substrate.

For protecting the YBCO thin film during patterning a second passivating PBCO layer with a thickness of 25 nm was deposited on top of the YBCO film. After the second PBCO-layer deposition the oxygen pressure was increased to 900 mbar, and the substrate temperature was ramped down to 550°C with a rate of 10°C/min. The temperature was kept constant at 550°C for 10 minutes for annealing of the obtained multilayer structure. Afterwards, the heater was ramped down to 400°C before switching off and cooling down exponentially to room temperature. The vacuum chamber was then pumped down to pressures below $5 \cdot 10^{-5}$ mbar and a 140 nm thick Au layer was grown *in-situ* using the same PLD technique. The energy density was increased to $\approx 3$ J/cm$^2$ for the Au deposition resulting in a deposition rate of 0.15 nm/s.

## III. DEVICE FABRICATION

### A. Microbridges

The as-deposited multi-layers were patterned by several electron-beam lithography and etching steps. The lengths of the detecting microbridges were defined using PMMA 950K resist with a thickness of 180 nm. To remove the Au from the active YBCO detecting bridge, a two-step etching process was used. First, the sample was etched with an Ar ion-milling process to remove the first $\approx 40$ nm of the Au layer. Since ion implantation and overheating of the YBCO layer destroy superconductivity, the remaining Au layer of 100 nm was removed by a wet-etching process in a $I_2KI$-solution for $\approx 30$ s. By this, the sample was etched isotropically until the microbridge was opened. A typical underetching of 0.5 µm was taken into account. The etching process was stopped using isopropanol.

To pattern the antenna and coplanar readout negative e-beam resist with a thickness of 350 nm was used. The final etching step was realized by Ar ion milling. Since PBCO and YBCO are both very hard materials (ceramic-like), the parameters for the ion milling were optimized towards high acceleration voltages. The acceleration voltage was increased to 520 V. During etching the sample was actively cooled to reduce oxygen loss in the YBCO layer due to overheating during the ion milling process.

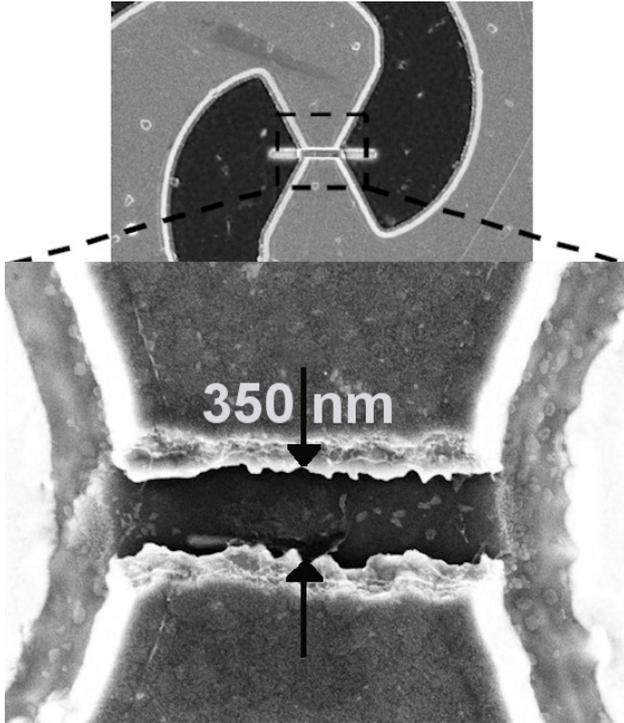

Fig. 2. SEM images of the center part of the antenna, with the YBCO detecting element of 350 nm length and 1.5 μm width.

A SEM image of a final device is shown in Fig. 1. The antenna was designed with CST Microwave Studios® [23] to encompass the typical THz frequency range of ANKA from 0.1 – 1 THz. Details concerning the antenna design are discussed elsewhere [25]. Typical dimensions of the YBCO microbridges at the center of the antenna for a 30 nm thick film were a width of 5 μm and a length of 2 μm.

*B. Nanobridges*

The main limitation for the reduction of the size of the detecting element to sub-micrometer dimensions, was the strong underetching of ≈ 0.5 μm in the $I_2KI$-solution during the patterning process as described above. To overcome this limitation the pre-etching in the Ar ion-milling system was extended to reduce the etching time in the $I_2KI$-solution and thus the underetching.

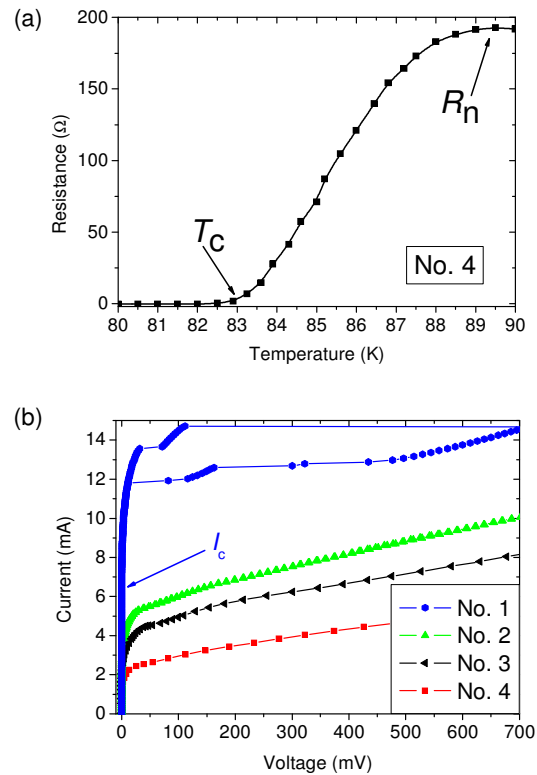

Fig. 3. (a) Temperature dependence of resistance for the smallest nanobridge with a length of 296 nm and a width of 910 nm (sample No. 4). (b) Current-voltage characteristic for the micro- and nanometer sized YBCO bridges measured at 77 K in liquid nitrogen.

For the extension of the ion-milling etching procedure a thicker resist was required. Thus, for patterning of the detecting elements a PMMA 950K resist with a thickness of 460 nm was used.

The samples were etched in the Ar ion-milling system to remove most of the Au layer (≈ 110 nm). The remaining Au layer in the detecting slits after ion-milling of ≈ 30 nm was then removed by wet etching using the $I_2KI$-solution. By this, the underetching could be reduced to less than 80 nm and detecting elements as short as 300 nm were successfully realized. Fig. 2 shows the center part of a final device with a YBCO nanometer-sized detector embedded in the Au antenna.

## IV. DC CHARACTERISATION

The detectors were characterized by measuring the temperature dependence of resistance and the current-voltage curves in liquid nitrogen. The temperature dependence of resistance for the smallest nanobridge with a length of 300 nm and a width of 900 nm is shown in Fig. 3(a). The zero-resistance critical temperature $T_c$ was determined to 82.9 K. This is a reduction of only ≈ 3 K compared to the micrometer-sized samples. The normal state resistance was determined right above the onset of the superconducting transition (see Fig. 3(a)) and amounted to 193 Ω for sample No. 4.

TABLE I
CHARACTERISTICS OF MICRO- AND NANOMETER SIZED YBCO THZ DETECTORS

| Sample No. | YBCO film thickness (nm) | Width (μm) | Length (μm) | Critical temperature (K) | Transition width (K) | Critical current density at 77 K (MA/cm²) | Normal state resistance (Ω) |
|---|---|---|---|---|---|---|---|
| 1 | 30 | 5.0 | 2.0 | 85.7 | 1.48 | 4.1 | 81 |
| 2 | 30 | 3.2 | 0.93 | 83.9 | 2.62 | 2.2 | 101 |
| 3 | 25 | 2.7 | 0.72 | 82.2 | 3.96 | 1.6 | 141 |
| 4 | 30 | 0.91 | 0.296 | 82.9 | 3.74 | 2.9 | 193 |

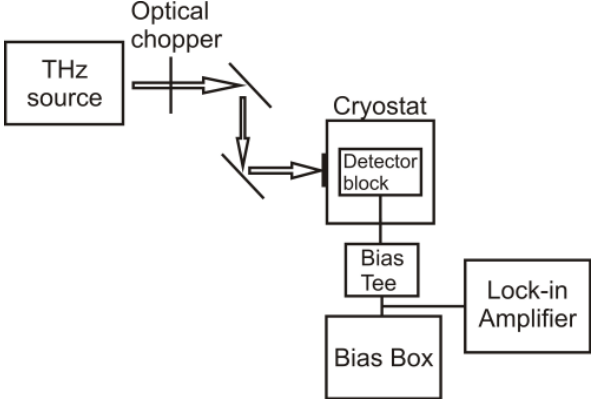

Fig. 4. Scheme of experimental setup to measure the THz response of YBCO thin film detectors.

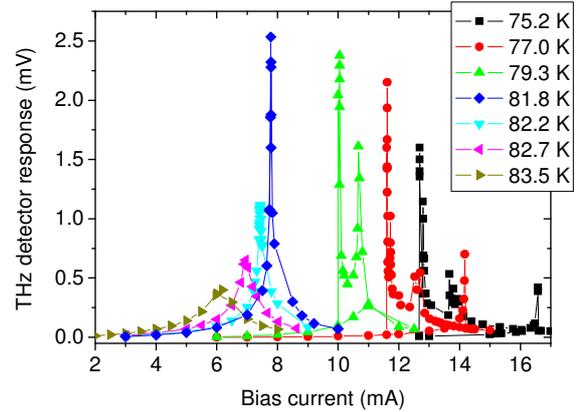

Fig. 5. Temperature and bias current dependence of the detector response at 0.65 THz for sample No. 1. The maximum response was achieved at 81.8 K.

The current-voltage characteristics of the samples were measured in liquid nitrogen and are displayed in Fig. 3(b). The critical current was determined right before the onset of the voltage drop caused by dissipative flux-flow in the bridges (indicated by the arrow for sample No. 1 in Fig. 3(b)). A decrease of the critical current with the reduction of the bridge width is observed as expected.

All details concerning the superconducting characteristics of the nano- and micrometer-sized YBCO bridges are summarized in Table I. The reduction of the critical temperature and the increase of the transition width with the reduction of the bridge width can be explained by out-diffusion of oxygen caused by damaged edges. This reduction of the critical temperature also explains the reduction of the critical current density for the smaller bridges (see Table 1).

## V. THz CHARACTERISATION

### A. Measurement setup

The setup for the characterization of our YBCO thin film detectors with THz radiation is shown in Fig. 4. As THz sources either continuous 0.65 THz radiation optically chopped at 15 Hz or the ANKA storage ring with pulsed emission (0.1 – 2 THz) were used. The THz radiation is guided by parabolic mirrors to the polyethylene window of the cryogen-free cryocooler. The radiation is then focused by a silicon lens, embedded in a copper detector block, to the detector chip which is rear-side illuminated. The detector block is mounted to the cold-finger of the cryocooler and connected via semi-rigid cables to a vacuum feed-through leading to a room-temperature bias-tee. The detector is biased via the bias-tee with a battery-driven source. For the continuous wave irradiation the detector signal is read out via a T-adapter at the bias line with a lock-in amplifier (see Fig. 4). For the pulsed excitation at ANKA the 33 GHz Agilent real-time oscilloscope (DSOX93204A) was directly connected to the bias-tee. Due to the high responsivity values of our nanometer-sized detectors no amplifier is required in the read-out.

### B. Responsivity measurements at CW THz radiation

For the determination of the maximum responsivity of the micro- and nanobridges, the detector response was measured in dependence on bias current and temperature. In Fig. 5 the typical detector response to 0.65 THz radiation for various temperatures and bias currents is shown.

The responsivity $S$ is defined as the ratio of the detector output signal $\Delta U$ to the absorbed radiation power $\Delta P$. For continuous wave excitation with the modulation frequency $\omega$ and the detector time constant $\tau$ where $\omega\tau \ll 1$ the frequency-dependence of the responsivity can be neglected and $S$ is calculated according to

$$S = \frac{\Delta U}{\Delta P} = \frac{dR/dT\, I_b}{G_{eff}\sqrt{1+(\omega\tau)^2}} = \frac{dR/dT\, I_b}{G_{eff}}\bigg|_{\omega\tau\ll 1} \quad (1)$$

with the temperature derivative of resistance $dR/dT$, the bias current $I_b$ and the effective thermal conductance $G_{eff}$ which is calculated according to $G_{eff} = G - dR/dT\, I_b^2$ where $G$ is the thermal conductance at zero applied current.

The radiation power level was measured with a THz power meter [24] to $P_{rad} = 80$ μW at the focal point of the second parabolic mirror. The maximum detector signal for the YBCO microbridge of 2.5 mV is achieved at 81.8 K and a bias current of 7.78 mA. Thus, the optical system responsivity amounts to 31 V/W.

To determine the responsivity of the YBCO bridges at 0.65 THz the system coupling efficiency has to be taken into account. The transmission of the HDPE window of the cryocooler was determined experimentally. It amounts to 80%. The other losses were calculated by numerical simulations. The transmission at the lens-air interface is calculated to 67%. The Gaussian beam coupling efficiency resulted in 42%, while the polarization coupling efficiency was calculated to 50%. The coupling between the antenna and the detecting element is dependent on the impedance of the sample. To avoid uncertainties resulting from poor knowledge of the rf impedance of our samples, the following approximation for the impedance will be used as discussed in [25]: $Z_d = \text{Re}(Z_d) + i\,\text{Im}(Z_d)$, with $\text{Re}(Z_d) \approx \text{Im}(Z_d) \approx R_n$. Taking into account the almost real impedance of the antenna of $Z_a = 65\,\Omega$, the coupling efficiency between antenna and detector can be calculated according to

$$\eta = 1 - |\underline{S_{11}}|^2 = 1 - \left|\frac{Z_d - Z_a^*}{Z_d + Z_a^*}\right|^2 \qquad (2)$$

resulting in an antenna-detector coupling efficiency of 75% and 48% for sample No. 1 and No. 4, respectively.

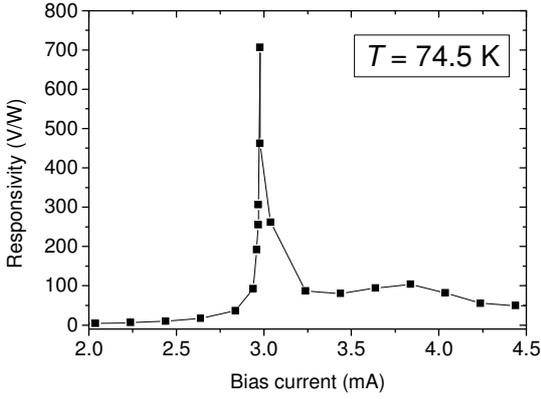

Fig. 6. Bias current dependence of the responsivity for sample No. 4 at 0.65 THz. The maximum responsivity was achieved for a bias current of 3 mA at 74.5 K.

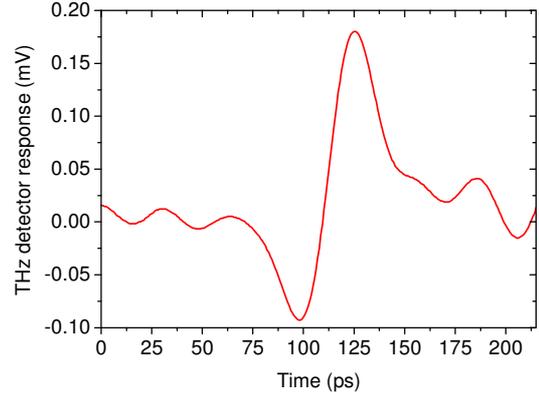

Fig. 7. Zero-bias detector response of sample No. 4 measured at 77 K. The pulse shape is in good agreement with simulations [28].

The coupling efficiency of the total system results in 8.4% and 5.4%, respectively, which is in good agreement with other published values [26]. Taking into account the system coupling efficiency this results in a maximum responsivity for sample No. 1 of 372 V/W.

For the nanobridge devices the responsivity could be increased by a factor of $\approx 2$. In Fig. 6 the THz detector responsivity as a function of the bias current for the smallest nanobridge (sample No. 4) is shown. The maximum value of 710 V/W was achieved at 74.5 K for a bias current of 3 mA.

The increase of the responsivity by a factor of 1.9 is well explained by equation (1). The temperature derivative, deduced from the experimental $R(T)$ curves, was a factor 3 larger for the microbridge than for the nanobridge which can be explained by the above mentioned broadening of transition. Also the bias current was 2.5 times larger for the microbridge. However, the reduction of the area of the nanobridge which is directly proportional to the thermal conductance lead to a decrease of the effective thermal conductance from 940 µW/K for the microbridge to only 63 µW/K for the nanobridge, respectively, thus explaining the increase in the responsivity for the nanobridge.

Further improvement of the responsivity values for our YBCO nanobridges is readily achievable by the increase of the YBCO film thickness resulting in steeper superconducting transitions and higher critical currents thus allowing to operate the detectors at higher bias currents.

### C. Zero-bias response at pulsed excitations

The detection of CSR THz pulses at the electron storage ring ANKA has already been demonstrated [15], [27]. The filling patterns of the storage ring with three consecutive trains as well as single pulses were successfully monitored. Single pulses as short as 6.8 ps (root mean square) were recorded [15].

For pulsed excitations in the THz frequency range the authors recently demonstrated that the detection mechanism is based on dissipative movement of vortices. The movement is caused by the Lorentz force of the excited rf current in the detector due to the electric field of the CSR pulse [25]. This explains why it is possible to measure detector responses even without any bias applied to the detector (which is not the case for a bolometric detector). The bias-free operation of the detectors clearly facilitates the experimental setup. Even more important is the fact that at zero bias the detector is able to follow the polarity of the exciting electrical field of the CSR pulse, thus allowing to measure the time-evolution of the emitted electrical field of the storage ring directly in the time-domain.

The high responsivities of our nanobridges allowed to measure at zero bias without any amplification. However, the determination of the responsivity to pulsed THz radiation is not possible since the exact pulse duration of the THz CSR radiation is unknown.

The zero-bias response of sample No. 4 measured at 77 K is shown in Fig. 7. The pulse shape with a negative and positive peak is in very good agreement with simulations regarding the time structure of the electric field from a Gaussian bunch for the electron storage ring ANKA [28] thus demonstrating that YBCO detectors are a powerful tool for the analysis of CSR emission.

### VI. CONCLUSION

Pulsed-laser deposition was used for the fabrication of 30 nm thick YBCO samples on sapphire. The patterning procedure was optimized for sub-micrometer dimensions and nanometer-sized devices as small as 300 nm x 900 nm were fabricated. YBCO micro- and nanobridges were characterized at THz frequencies with continuous wave radiation at 0.65 THz reaching responsivity values of more than 700 V/W. This allows embedding our thin-film nanobridges in an ultra-fast readout system without amplifier. Thus, it was possible to measure a clear detector response to the picosecond CSR pulses emitted at ANKA even without any bias. The expected pulse shape of the electrical field with negative and positive peak was successfully resolved. Our YBCO thin-film detectors are, therefore, suitable for monitoring CSR THz pulses opening new routes for the analysis and understanding of CSR generation and dynamics.

ACKNOWLEDGMENT

The authors want to thank M. Kohler und M. Stocklas from Agilent Technologies for supplying a 33 GHz real-time oscilloscope for the measurements at ANKA. This work was supported by the German Federal Ministry of Education and Research (Grant Nos. 05K10VKD, 05K10KTC).